\documentclass[
 aps,
 prd,
twocolumn,
superscriptaddress,
frontmatterverbose, 
showpacs,preprintnumbers,
 nofootinbib,
 amsmath,
 amssymb,
floatfix,
latexsym,array,enumerate,letter,
]{revtex4-1}

\usepackage[utf8]{inputenc}
\usepackage[english]{babel}
\usepackage{multirow}
\allowdisplaybreaks
\usepackage{graphicx,color}
\usepackage[colorlinks=true,citecolor=blue,linkcolor=blue,urlcolor=blue]{hyperref}
\usepackage{comment}
\usepackage{appendix}
\usepackage{caption}
\usepackage{ulem}

\usepackage{subcaption}
\usepackage{footmisc}
\usepackage{bm}
\usepackage{dcolumn}
\usepackage{gensymb}
 \usepackage{float} 
\restylefloat{table}
\usepackage{multirow}
 \usepackage{hyperref}
 \usepackage{array}

\hypersetup{
    colorlinks,
    citecolor=black,
    filecolor=black,
    linkcolor=blue,
    urlcolor=blue
}

\newcounter{infer}[section]
\renewcommand*{\theinfer}{}

\newcolumntype{P}[1]{>{\centering\arraybackslash}p{#1}}
 \newcommand{\be}{\begin{equation}}
\newcommand{\ee}{\end{equation}}
\newcommand{\beq}{\begin{equation}}
\newcommand{\eeq}{\end{equation}}
\newcommand{\bea}{\begin{eqnarray}}
\newcommand{\eea}{\end{eqnarray}}

\def \W  {\mathcal{W}}
\def \V  {\mathcal{V}}

\begin{document}

\title{Self-bound hybrid stars with strong phase transitions can relieve major compact star observation tensions}
\author{Chen Zhang}
\email{zhangvchen@tongji.edu.cn (Corresponding Author)}
\affiliation{School of Physics Science and Engineering, Tongji University, Shanghai 200092, China}
\affiliation{The HKUST Jockey Club Institute for Advanced Study,
The Hong Kong University of Science and Technology, Hong Kong, P.R. China}

\author{Juan M. Z. Pretel}
\email{juanzarate@cbpf.br}
 \affiliation{
 Centro Brasileiro de Pesquisas F{\'i}sicas, Rua Dr.~Xavier Sigaud, 150 URCA, Rio de Janeiro CEP 22290-180, RJ, Brazil
}

\author{Renxin Xu}
\email{r.x.xu@pku.edu.cn}
\affiliation{Department of Astronomy, School of Physics, Peking University, Beijing 100871, China}
\affiliation{Kavli Institute for Astronomy and Astrophysics, Peking University, Beijing 100871, China}

\begin{abstract}

Some recent pulsar observations cannot naturally fit into the conventional picture of neutron stars: the compact objects associated with HESS J1731-347 and XTE J1814-338 have too small radii in the low-mass regime, while the secondary component of GW190814 is too massive for neutron stars to be compatible with constraints from the GW170817 event. In this study, we demonstrate that all these anomalous observations and tensions, together with other conventional ones such as recent NICER observations of PSR J0740+6620, J0030+0451, and PSR J0437-4715, can be naturally explained simultaneously by a new general type of self-bound hybrid stars with large density discontinuities, and thus are radially stable in either the slow or rapid phase transition context. As a proof of concept, we use hybrid quark stars, inverted hybrid stars, and hybrid strangeon stars as benchmark examples to explicitly demonstrate the advantage and feasibility of self-bound hybrid stars with strong phase transitions in relieving all tensions related to compact stars' masses, radii, and tidal deformabilities.
\end{abstract}
\maketitle
\section{Introduction}

Phase transitions within compact stars are a long-studied possibility~\cite{doi:10.1142/S0218301318300084,orsaria2019phase,heiselberg1999phase,glendenning2001phase,yang2008influence,reddy2000first,glendenning1992first,burgio2002hadron}, which could provide a unique opportunity to understand the fundamental strong interaction between quarks, in both the perturbative and the nonperturbative regimes. More and more research indicates that quark matter (QM), a state composed of deconfined free-flowing quarks, can possibly exist inside the neutron star core from hadron-quark phase transitions, forming so-called hybrid stars~\cite[e.g.,][]{Alford:2004pf,Alford:2013aca}. 

It is possible that QM is absolutely stable (i.e., energy per baryon number $E/A<930$ MeV) even at zero pressure, either in the form of strange quark matter  (SQM)~\cite{Bodmer:1971we,Witten:1984rs,Terazawa:1979hq,Farhi:1984qu} or up-down quark 
 matter ($ud$QM)~\cite{Holdom:2017gdc,Wang:2021byk,Xia:2022tvx, Bai:2024muo,Bai:2024amm}. It can constitute the entire star~\cite{Haensel:1986qb,Alcock:1986hz, Xu:1999bw,Yang:2023haz,Zhang:2019mqb, zhao2019current, Ren:2020tll,Cao:2020zxi,Yuan:2022dxb,Restrepo:2022wqn,Xia:2020byy,Zhang:2020jmb,Pretel:2024pem,Zhou:2024syq,Wang:2024xon,Xie:2025sth,Chen:2025ppr,Miao:2021nuq,Miao:2024qik}, forming the so-called quark star that is self-bound and featuring a high-density surface. Possible phase transitions can also occur inside the quark stars, leading to self-bound hybrid stars such as hybrid quark stars (HybQSs) from a QM-QM (different QM phases) transition~\cite{HybQS}, or inverted hybrid stars (IHSs) from a speculated quark-hadron phase transition~\cite{Zhang:2022pse,Zhang:2023zth,Negreiros:2024cvr}.
Clustered SQM, or in another name Strangeon Matter (SM), like strange quark matter, comprises nearly equal numbers of $u,d,s$ quarks~\cite{Xu:2003xe, Lai:2009cn, Lai:2017ney} as favoured by the approximate SU(3) flavor symmetry, and could also be absolutely stable, forming a solid state with localized quark clusters (strangeons) due to their large mass and strong interaction. Strangeon stars~\cite{Xu:2003xe,Lai:2017ney,Lai:2009cn,Miao:2020cqj,Gao:2021uus,Zhang:2023mzb}, composed of SM and thus self-bound, have a stiff equation of state (EOS) and high compactness that helps address various astrophysical observations~\cite{Zhang:2023mzb,Li:2023tng,Li:2024hzt, Wang:2024opz,Yuan:2024hge}. A transition from SM to SQM is possible~\cite{Miao:2020cqj}, forming the so-called Hybrid Strangeon Stars (HybSSs) with a strangeon crust and an SQM core~\cite{Zhang:2023szb}.

The stability of compact stars is generally determined by the fundamental ($f$) mode frequency of radial oscillations, whose squared eigenfrequencies' ($\omega_0^2$) positive signs indicate the star's stability against small adiabatic perturbations. Studies of radial pulsations in hybrid stars are typically categorized into two types: slow and rapid transitions, depending on the phase-transition timescale relative to the oscillation timescale. For rapid transitions, the region where $\omega_0^2\geq0$ coincides with the conventional stability condition $\partial M/ \partial P_c\geq0$ (i.e., star mass $M$ increases with increasing center pressure $P_c$), implying that the zero fundamental eigenfrequency point coincides with the maximum mass point~\cite{Glendenning:1997wn,Pereira:2017rmp,Pretel023524}. However, this coincidence does not hold for the slow-transition case, where an extended region of stability develops after a sharp phase transition even with $\partial M/ \partial P_c<0$, corresponding to the so-called slow stable hybrid stars~\cite{Pereira:2017rmp,Lugones:2021zsg,Lugones:2021bkm,Goncalves:2022phg,Rau:2022ofy}, which can be formed from early phase transitions triggered by quantum or thermal nucleations in realistic astrophysical scenarios such as protoneutron stars or the postmerger hot stars~\cite{Lugones:2021bkm}. The physics that determines whether the transition is the slow or rapid type has large uncertainties due to the unperturbative nature of strong dynamics. The interface surface tension, which slows down the transition, suffers from the notorious model-dependence. However, we can expect a slow phase transition if the flavor composition of the two phases is different, since a high-order weak interaction process would be needed.

In 2022, Doroshenko \textit{et al.}~\cite{Doroshenko-2022} reported puzzlingly low mass-radius values ($M=0.77^{+0.20}_{-0.17}\, M_\odot$, $R=10.4^{+0.86}_{-0.78}$ km at $1\sigma$) for the central compact object in the supernova remnant HESS~J1731-347, which prompted extensive studies proposing several explanations such as hybrid star model with early phase transition~\cite{Tsaloukidis:2022rus,Sagun:2023rzp,Laskos-Patkos:2023tlr, Laskos-Patkos:2024otk, Mariani-2024,Gao:2024chh,Pal:2025skz,Hong:2024srz}, strange quark star model~\cite{Di_Clemente_2024,Horvath:2023uwl,Oikonomou_2023,Rather:2023tly,Gholami:2024ety,Issifu_2025,Kourmpetis:2025zjz}, a soft nuclear EOS neutron star~\cite{Brodie:2023pjw,Huang:2023,Li:2023vso,Kubis:2023gxa}, and mixed dark matter models~\cite{Hong:2024sey,Routaray:2023txs}. Recently, Kini \textit{et al.}~\cite{Kini:2024ggu} reported similarly puzzling ultracompact parameters for XTE~J1814-338 ($M=1.21^{+0.05}_{-0.05}\, M_\odot$, $R=7.0^{+0.4}_{-0.4}\, \rm km$ at $1\sigma$), with proposed explanations including a hybrid star branch~\cite{Laskos-Patkos:2024fdp, Zhou:2025uim}, a bosonic star with nuclear core~\cite{Pitz:2024xvh}, a dark matter admixed neutron star \cite{LOPES2025101922} and a strange star admixed with dark matter~\cite{Yang:2024ycl,Lopes:2025jyz,Yang:2025way}.  Nevertheless, it is very difficult to reconcile both puzzling objects at the same time~\cite{Veselsky:2024bnf,Laskos-Patkos:2024fdp,Kourmpetis:2025zjz}. In addtion,  a binary merger event GW190814 was reported~\cite{abbott2020gw190814}, featuring a primary black hole with mass $23.2^{+1.1}_{-1.0}\, M_{\odot}$, and a secondary companion of $2.59^{+0.08}_{-0.09}\, M_{\odot}$, which is much larger than the upper bound $M_{\rm TOV}\lesssim 2.3 M_{\odot}$ of the maximum mass for a nonrotating neutron star,
set by various analyses of GW170817~\cite{Margalit_2017,rezzolla2018using,ruiz2018gw170817,shibata2019constraint}. To the best of our knowledge, no previous studies have managed to build models that could accommodate all these anomalous observations at the same time.

Note that the mass–radius inferences for HESS J1731-347 and XTE J1814-338 should be interpreted with caution. For HESS J1731-347, the uniform carbon atmosphere model adopted in Ref.~\cite{Doroshenko-2022} is supported by the absence of pulsations and the observed blackbody-like spectra, with alternative atmosphere models explored but found less consistent with the data. For XTE J1814-338, the single-hotspot model of Ref.~\cite{Laskos-Patkos:2024fdp} successfully reproduces the fundamental fractional amplitudes and incorporates improvements over earlier analyses, though some discrepancies remain at higher harmonics, motivating future refinements with multi-hotspot models. 

In the absence of a comprehensive model capable of describing all observational mass-radius and tidal deformability measurements, this work examines slow stable self-bound hybrid stars with quite promising results. The general finding that self-bound stars like strange stars can saturate HESS~J1731-347 easily~\cite{Doroshenko-2022}, and the slow stable hybrid stars benefit to meet XTE J1814-338~\cite{Laskos-Patkos:2024fdp}, naturally motivates us to propose slow stable self-bound hybrid stars that combine these two advantages.  As we will explicitly show in the main text, the new branch that gives rise to the slow stable self-bound hybrid stars helps explain all these anomalous objects at the same time, while also fulfilling other astrophysical observations such as GW170817~\cite{LIGOScientific:2017vwq}, recent NICER analysis of PSR J0030+0451~\cite{Miller:2019cac,Riley:2019yda}, PSR J0740+6620~\cite{miller2021radius,riley2021nicer}, and the more recent PSR J0437-4715~\cite{choudhury2024nicer,rutherford2024constraining}.

This paper is organized as follows: first, we introduce models for the EOSs of three types of self-bound hybrid stars: HybQS, HybSS, and IHS. Then, we consider the mass-radius relations, radial oscillations, and tidal deformabilities with varying parameters, where we look for radially stable solutions that satisfy all the notable astrophysical constraints, particularly aiming to explain HESS~J1731-347 and XTE~J1814-338 at the same time. 
We employ the Maxwell construction for phase transitions and geometric units ($G=c=1$) throughout this paper. 

\section{Model of Self-bound hybrid stars}
To model the HybQS, we use the conventional MIT bag model EOS
\be
P=\frac{1}{3}(\rho-4B), \quad P<P_{\rm trans}
\label{bag}
\ee
for the quark matter composing the crust before the transition pressure $P_{\rm trans}$, with $B$ being the effective bag constant characterizing the QCD vacuum difference inside and outside the quark matter bag. We then adopt the constant-sound-speed (CSS) parametrization~\cite{Alford:2013aca} for the quark matter core with the EOS
\be
P=c_s^2 (\rho-\rho_{\rm trans}-\Delta{\rho})+P_{\rm trans}, \quad P>P_{\rm trans}
\label{CSS}
\ee
for the core quark matter after $P_{\rm trans}$, with $c_s^2$ denoting the sound speed squared, $\rho_{\rm trans}$ denoting the energy density at $P_{\rm trans}$, and $\Delta \rho$ as the energy density discontinuity of the phase transition interface. Therefore, we have four independent parameters  ($B$, $P_{\rm trans}$, $\Delta\rho$, $c_s^2$) for this HybQS construction.

To model HybSSs, the EOS of strangeon matter~\cite{Xu:2003xe, Lai:2017ney} located at the outer layer ($P<P_{\rm trans}$) can be written into the following form~\cite{Zhang:2023mzb,Zhang:2023szb}
\be
\begin{aligned}
\frac{P}{ n_s}&=\frac{2 \,a}{9}  \tilde{\epsilon}  \left(\frac{1}{9  }\bar{n}^5 -\bar{n}^3\right),\\
\frac{\rho}{ n_s}&= \frac{a}{9}  \tilde{\epsilon}  \left(\frac{1}{18  } \bar{n}^5 - \bar{n}^3\right)+ m_q \bar{n}, 
\end{aligned}
\label{EOS_strangeon}
\ee
where $a=A_6^2/A_{12}=8.4^2/6.2\approx11.38$, $m_{q}$ is the average constituent quark mass, and $n$ is the number density of strangeons with $n_s$ being the surface baryon number density. The normalized quantities  $\bar{n}=N_q \,n / n_s$  and $\tilde{\epsilon}=\epsilon/N_q$ with $\epsilon$ describing the depth of the interaction potential between strangeons, and $N_{\rm q}$ being the number of quarks in a strangeon. A larger $\epsilon$ will then indicate a larger repulsive force at short range and maps to a stiffer EOS.  The CSS parameterization (Eq.~\ref{CSS}) is then used to model the quark matter core $(P>P_{\rm trans})$. Thus, we have five independent parameters ($\tilde{\epsilon}$, $n_s$, $P_{\rm trans}$, $\Delta\rho$, $c_s^2$) for this HybSS construction.

To model IHSs~\cite{Zhang:2022pse,Zhang:2023zth,Negreiros:2024cvr}, we use the simple MIT bag model Eq.~(\ref{bag})
for the quark matter crust like the hybrid quark star model, and with the $e$-type polytrope EOS\footnote{Note that the polytope EOS can have different forms depending on whether the directly related variable is energy density ($e$-type) or rest mass density ($\mu$-type). For details, see discussions below Eq. (22) of Ref.~\cite{Damour:2009vw}. } ~\cite{Damour:2009vw} 
\be
P=K\rho^\gamma, \quad P>P_{\rm trans}
\ee
for the hadronic core. The constant pressure at the interface of the two matter phases yields 
\be
K=P_{\rm trans} (3 P_{\rm trans}+ \Delta{\rho}+4 B)^{-\gamma}.
\ee
Therefore, in this IHS model, we have four independent parameters ($B$, $P_{\rm trans}$,  $\Delta\rho$, $\gamma$). Due to the uncertainties associated with hadronic EOS, one can analyze the polytrope EOS of different parameter sets at different density regimes in a piecewise manner~\cite{hebeler2013equation,Huang:2025vfl}. 

The $E/A <$ 930 MeV constraint needs to be satisfied by the outer layer of these hybrid objects\footnote{The core composition can always be more stable after an arbitrary $P_{\rm trans}$ by tuning the extra parameters in the chemical potentials but not in EOS.}. For the noninteracting bag model of quark matter that we used to construct the outer layer of hybrid quark star and inverted hybrid stars, by taking the noninteracting limit ($a_4\to1$, $\bar{\lambda}\to 0$) in Eq.~(11) of Ref.~[31], the $E/A$ takes the form of 
 $$E/A=\frac{3\sqrt{2\pi}}{\xi_4^{1/4}} B^{1/4},$$
 where $\xi_4=3$ for three-flavor quark matter (SQM), and $\xi_4=( \left(\frac{1}{3}\right)^{\frac{4}{3}}+ \left(\frac{2}{3}\right)^{\frac{4}{3}})^{-3}\approx 1.86$ for two flavor quark matter ($ud$QM). Thus, the $E/A<930$ MeV constraint translates to an upper bound on the bag constant:
 $B<(\frac{\xi_4^{1/4}}{3\sqrt{2\pi}}930 \text{ MeV})^4$, yielding $B\lesssim(162.8 \text{ MeV})^4\approx91\, \rm MeV/fm^3$ for SQM~\footnote{Note that in the conventional absolutely-stable SQM picture, the $ud$QM is unstable, which translates to a lower bound $B\gtrsim 56.8\, \rm MeV/fm^3$, but this lower bound can be relaxed significantly by a smaller $a_4$ (i.e., large pQCD corrections), as can be seen from the $\bar{\lambda}\to 0$ limit of Eq.~(11) in Ref.~\cite{Zhang:2020jmb}, without affecting the EOS (since $a_4$ does not enter the EOS in this $\bar{\lambda}\to 0$ limit).}, and $B\lesssim (144.38 \text{ MeV})^4\approx 56.8\, \rm MeV/fm^3$ for $ud$QM.

For strangeon matter that composes the outer layer of a hybrid strangeon star, the $E/A$ takes the form of 
\be
E/A=3m_q-\epsilon/N_q
\ee
with $3m_q=930$ MeV~\cite{Zhang:2023mzb}, so that the parameter space always satisfies $E/A< 930$ MeV for positive $\epsilon$. 

\section{Stellar properties and stability}
The static spherically symmetric background spacetime has the following line element
\begin{equation}
ds^{2}=-e^{ 2\Phi(r)} dt^{2} + e^{ 2\Psi(r)} dr^{2} + r^{2}(d\theta^{2}+\sin^{2}{\theta}d\phi^{2}),
\end{equation}
where $e^{-2\Psi(r)} = 1 - 2m(r)/r$, with $m(r)$ being a mass function within the hybrid star.
To obtain the configuration of relativistic stellar structure, we incorporate the combined EOS described above into the so-called Tolman-Oppenheimer-Volkov (TOV) equations~\cite{Oppenheimer:1939ne,Tolman:1939jz}
 \bea
 \begin{aligned}
{dP(r)\over dr}&=-{\left[m(r)+4\pi r^3P(r)\right]\left[\rho(r)+P(r)\right]\over r(r-2m(r))}\,,\,\,\\
{dm(r)\over dr}&=4\pi\rho(r)r^2\,, \\
\frac{d\Phi(r)}{dr} &= - \frac{1}{\rho(r)+ P(r)} \frac{dP(r)}{dr}, 
\end{aligned}
\label{eq:tov}
\eea
with the boundary conditions $
 \rho(0) = \rho_{\mathrm{c}},\, \Phi(R) = -\Psi(R)\,,
  $
where the star's radius $R$ and mass $M$ are determined by the conditions $P(R)=0$ and $m(R)=M$, respectively.


To investigate the radial stability of our hybrid stellar models, we assume that a fluid element is displaced from its equilibrium position $r$ to the perturbed position $r+\Delta r$, where such a perturbation has a harmonic time dependence $e^{i\omega t}$.  
The equations for solving infinitesimal radial oscillations $\xi=\Delta r/r$ and the corresponding pressure perturbations $\Delta P$ of a spherical object are~\cite{chanmugam1977radial, vath1992radial, Pereira:2017rmp,Gondek1997,Vasquez2010,Pretel2020MNRAS}
\begin{eqnarray}
\frac{d\xi}{dr}&=&\V(r)\xi+\W(r)\Delta P, \label{eqXi} \\
\frac{d\Delta P}{dr}&=& X(r) \xi + Y(r)  \Delta P, \label{eqDP}
\end{eqnarray}
with the coefficients given by
%
\be
\begin{aligned}
\V(r) &= -\frac{3}{r}-\frac{dP}{dr}\frac{1}{(P+\rho)}, \qquad
\W(r) = -\frac{1}{r}\frac{1}{\Gamma P},  \\
X(r) &= \omega^{2}e^{2\Psi-2\Phi}(P+\rho)r-4\frac{dP}{dr} \\
 &  + \bigg(\frac{dP}{dr}\bigg)^{2}\frac{r}{(P+\rho)}-8\pi e^{2\Psi}(P+\rho) Pr , \\
Y(r) &= \frac{dP}{dr}\frac{1}{(P+\rho)}- 4\pi(P+\rho)r e^{2\Psi}, \label{ecuacionparaP}
\end{aligned}
\ee
where $\Gamma=\frac{n}{P}\frac{dP}{dn}=\frac{\rho+P}{P}\frac{dP}{d\rho}$ is the adiabatic index. The initial conditions are $
(\Delta P)_{r=0}=-3(\xi \Gamma P)_{r=0}$ and $\xi(0)=1$. We obtain the squared eigenfrequencies $\omega^2$ by solving Eq.~(\ref{eqXi}) and Eq.~(\ref{eqDP}) using the shooting method, so that the boundary condition
$(\Delta P)_{r=R}=0
$
is satisfied at the surface. The fundamental zero mode that determines the radial stability is the lowest-lying one, the frequency of which is commonly denoted as $\omega_0$. For rapid transitions, the matching conditions at the interface are
$[\Delta P]^+_-=0$ and $\left[\xi -\frac{\Delta P}{r P'} \right]^+_-=0$, where $[x]^+_-\equiv x^+-x^-$ and primes on the variables denote the partial derivatives with respect to $r$. In this context, the stable region determined from these radial oscillation calculations generally coincides with the simple criterion $\partial M/\partial P_c>0$~\cite{Glendenning:1997wn,Pereira:2017rmp}.
For slow transitions, in which we are most interested, the matching conditions across the interface of the two matter phases are $[\xi]^+_- = 0$ and $[\Delta P]^+_-=0$. The classical stability criterion $\partial M/\partial P_c>0$ generally fails in this context, and it is necessary to use the signs of squared eigenfrequencies to determine the stability~\cite{Pereira:2017rmp,Lugones:2021zsg,Pretel023524,Goncalves:2022phg}.

For comparisons with gravitational wave observations, the dimensionless tidal deformability $\Lambda=2k_2/(3C^5)$ can be computed, where $C=M/R$ is the compactness and $k_2$ is the Love number characterizing the stellar response to external perturbations~\cite{hinderer2008tidal,hinderer2010tidal,postnikov2010tidal}.
The Love number $k_2$ is determined by solving a differential equation for $y(r)$ \cite{postnikov2010tidal} concurrently with the TOV equation Eq.~(\ref{eq:tov}), using the boundary condition $y(0)=2$. For self-bound hybrid stars, the matching condition $y(r_{d}^+) - y(r_{d}^-) = -4\pi r_{d}^3 \Delta \rho_d /(m(r_{d})+4\pi r_{d}^3 P(r_{d}))$ must be applied at $r_d$ (i.e., at both the core radius where $r_{\rm core}=r (P=P_{\rm trans})$ and the star surface $R$), where an energy density jump $\Delta \rho_d$ occurs~\cite{damour2009relativistic,takatsy2020comment}.
\section{RESULTS AND Discussions}
We aim to explain the puzzling XTE J1814-338 and HESS J1731-347 together with other astrophysical observations utilizing the slow stable branch of self-bound hybrid stars. There are two possible scenarios:
\begin{itemize}
    \item Scenario 1:  GW170817's $\Lambda_{1.4 M_{\odot}}<800$ constraint \footnote{Note that other constraints (such as the lower bounds of tidal deformability) from analyses of GW170817 commonly assume neutron star EOS~\cite{LIGOScientific:2018cki,LIGOScientific:2018hze, radice2018gw170817,kiuchi2019revisiting}, so that they do not apply to our study of self-bound stars. For details, see related discussions in Refs.~\cite{Zhang:2019mqb, Miao:2021nuq}.} needs to be met by the nonhybrid branch (i.e., by the bare quark stars of HybQS and IHS cases, and by the bare strangeon stars of HybSS case).
    \item Scenario 2: GW170817's constraint is only met by the hybrid branch (i.e., after $P_{\rm trans}$). 
\end{itemize}
In all scenarios, the HESS J1731-347 constraint can be easily met by all the nonhybrid self-bound branches, while the XTE J1814-338 is conveniently met by the (slow stable) hybrid branches in a viable EOS parameter space. 

Studies~\cite{LIGOScientific:2018hze,Radice:2017lry,Most:2018hfd,De:2018uhw,Burgio:2018yix,shibata2019constraint} have shown that GW170817's tidal deformability constraints $\Lambda_{1.4 M_{\odot}}<800$ exclude very stiff EOSs and large radii around $1.4\, M_\odot$ ($R_{1.4M_{\odot}}$), constraining possible maximum mass $M_{\rm TOV}$ well below that of the massive object ($2.6\,M_\odot$) observed in GW190814. Interestingly, Scenario 2 can relax $R_{1.4M_{\odot}}$ and $M_{\rm TOV}$ to large values, and reconcile the tension between the gravitational wave events GW170817 and GW190814 by leaving GW170817's constraints satisfied only by the hybrid branch, as we will explicitly show later.

\begin{figure*}[htb]
\centering
\includegraphics[width=8cm]{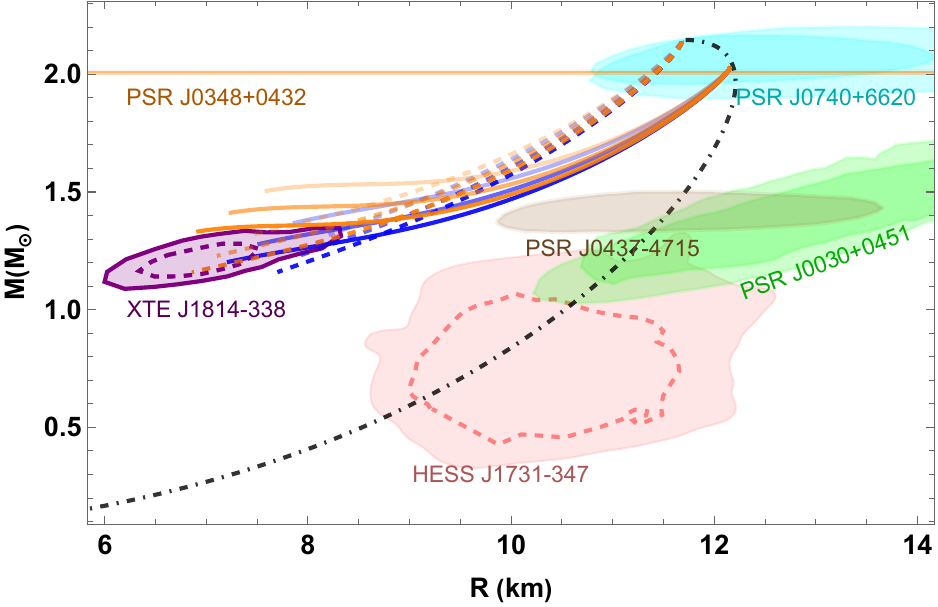}   
\includegraphics[width=8cm]{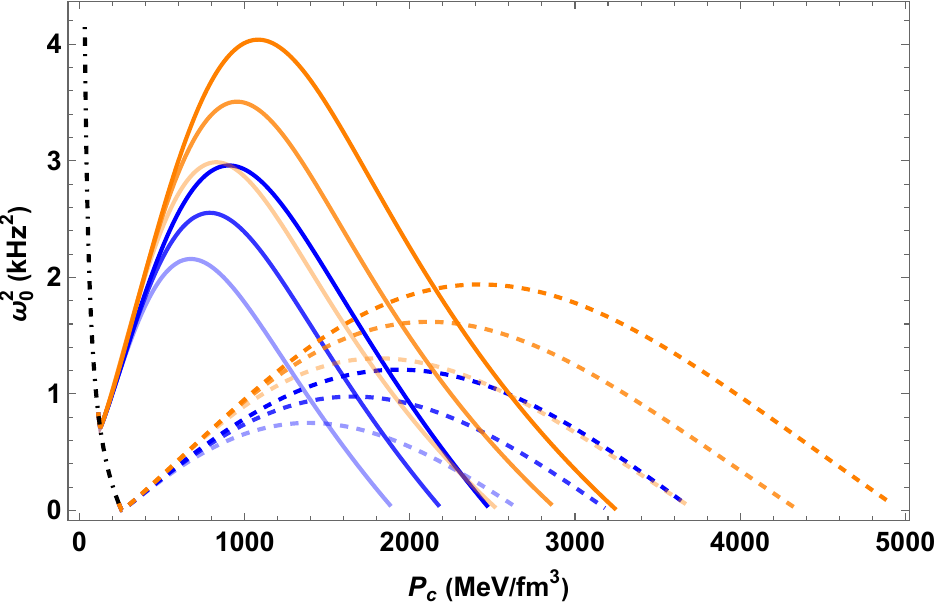}   
\includegraphics[width=8cm]{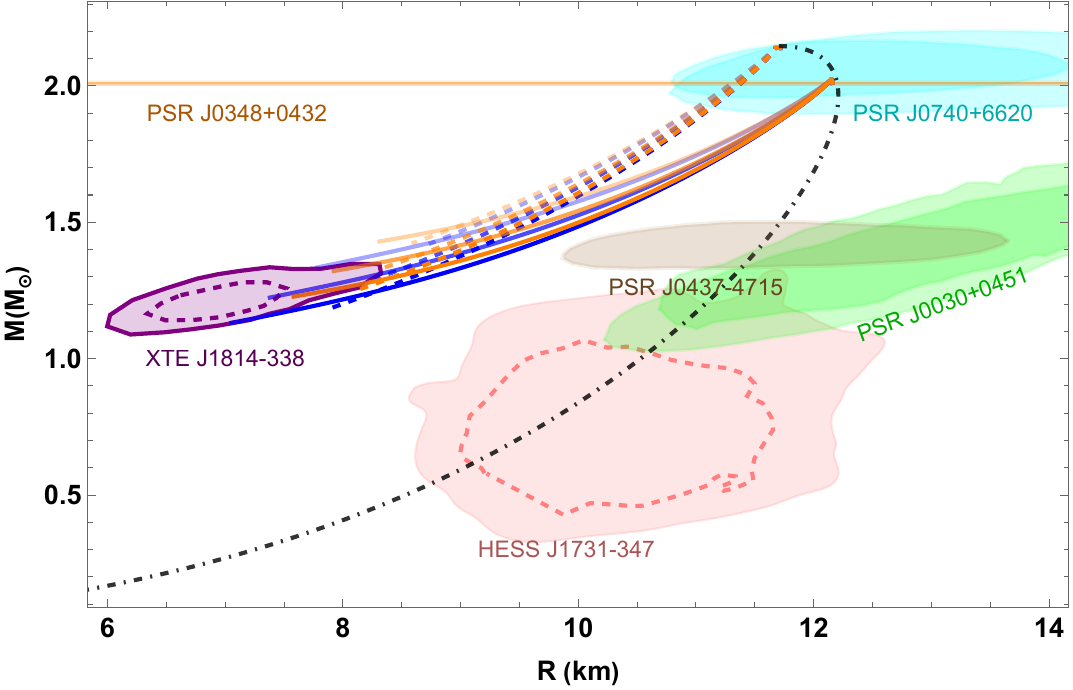}   
\includegraphics[width=8cm]{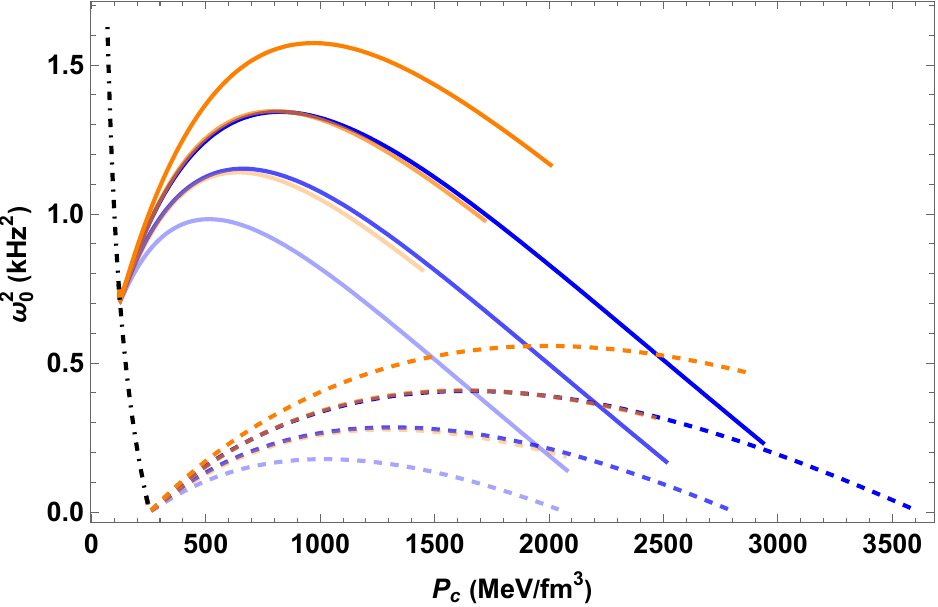}  
\includegraphics[width=8cm]{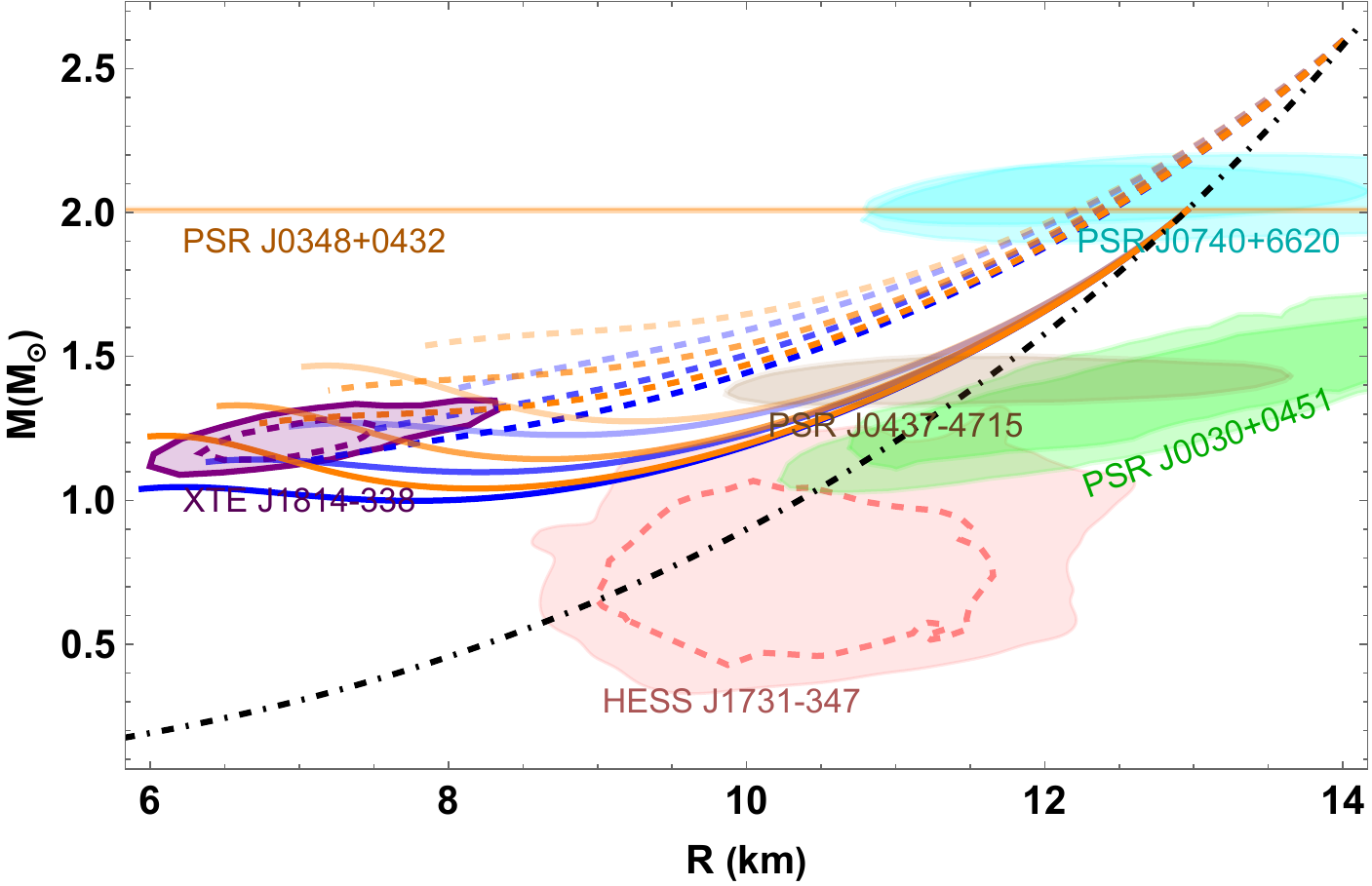}   
\includegraphics[width=8cm]{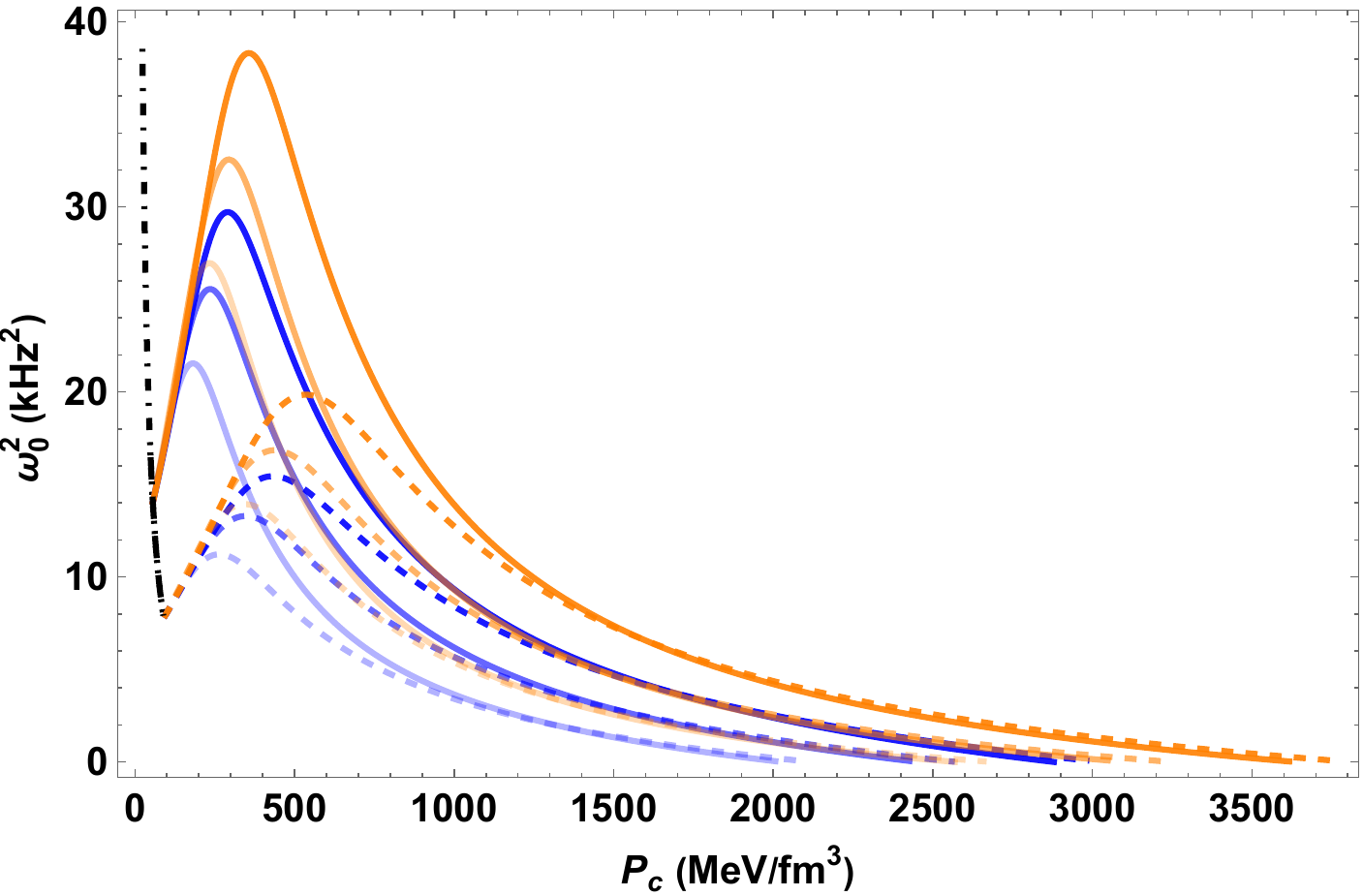}  
 \caption{Mass-radius relations (left column) and squared eigenfrequencies of fundamental radial oscillation mode versus center pressure (right column) for slow stable hybrid quark stars (first row), inverted hybrid stars (second row), and hybrid strangeon stars (last row), all in Scenario 1 where the nonhybrid branches (black dot-dashed) satisfy $\Lambda_{1.4M_\odot}<800$ (GW170817).
 For the first row, $B=50.3\rm \, MeV/fm^3$ for the quark matter crust. Blue ($c_s^2=0.6$) and orange ($c_s^2=1$) curves darken with larger $\Delta \rho/\rho_{\rm trans}$, sampling $(1.5,1.8,2.1)$ respectively. 
In the second row, $B=50.3\, \rm MeV/fm^3$ for the quark matter crust. Blue ($\gamma=2.1$) and orange ($\gamma=2.3$) curves darken with increasing $\Delta\rho/\rho_{\rm trans}$, sampling (1, 1.2, 1.4) respectively. 
For the third row, $\epsilon/N_q=80/9$ MeV for the strangeon matter crust, fixing $n_s=0.3\,\rm fm^{-3}$. Blue ($c_s^2=0.6$) and orange ($c_s^2=1$) curves darken with larger $\Delta \rho/\rho_{\rm trans}$, sampling $(4,5,6)$ respectively. 
For all rows, solid and dashed line styles denote cases where the transition point is around $2\, M_{\odot}$  and a larger selected mass, where $P_{\rm trans}=125\rm\, MeV/fm^3$ and $255\rm\, MeV/fm^3$ for HybQSs, $P_{\rm trans}=55\rm\, MeV/fm^3$ and $135\rm\, MeV/fm^3$ for IHSs, $P_{\rm trans}=68\rm\, MeV/fm^3$ and $90\rm\, MeV/fm^3$ for HybSSs,  respectively.  
 Color bands show astrophysical constraints. All curves terminate where either radial stability ($\omega_0^2 \geq 0$) or causality ($c_s^2 \leq 1$) is violated.
 }
 \label{fig_S1}
\end{figure*}

First, for Scenario 1, we choose the EOS parameters of the out-layer matter (i.e., bag constant $B$ of quark matter for HybQSs and IHSs, ($\epsilon/N_q,n_s$) of strangeon matter for HybSSs) so that GW170817's $\Lambda_{1.4M_\odot}<800$ bound is satisfied by the nonhybrid branch~\cite{Zhang:2019mqb,Zhang:2023mzb} before the phase transition (i.e., $P_c<P_{\rm trans}$). We display the results of Scenario 1 in Fig.~\ref{fig_S1}. The nonhybrid branch is naturally identified as the compact object associated with HESS J1731-347, GW170817, and all the NICER-related PSR J0030+0451, PSR J0740+6620 and the very recent PSR J0437-4715. At the more massive region above $2\,M_{\odot}$, a hybrid branch is developed from a sharp first-order phase transition, where the large densities discontinuities at the phase transition interface will render $\partial M/\partial P_c<0$, however, still be radially stable in the context of slow phase transition even until the XTE J1814-338's constrained $M$-$R$ region. Note that for the case of inverted hybrid quark stars (second row of Fig.~\ref{fig_S1}), some of the $\omega_0^2(P_c)$ curves that are not terminated at zero frequency because they violate causality ($c_s^2>1$) earlier before reaching zero $\omega_0^2$, due to a combined effect of their large polytrope index for the core EOS and the large center pressure extended.

We can see that all three types of self-bound stars can have viable parameter space for their slow stable branch (colored curves) to satisfy the constraints of XTE J1914-338, while the nonhybrid star branch (dot-dashed black curves) before the phase transition meets all other constraints without the need for much fine-tuning. In addition, it can be seen that the slow stable branch of all hybrid star types manifests the following common features:
\begin{itemize}
    \item $\omega_0^2$ increases after the phase transition, with a larger slope and eigenfrequency peak for a smaller $P_{\rm trans}$ (dashed) or a larger $\Delta\rho/{\rho_{\rm trans}}$ (darker) or stiffer core EOS (orange).
    \item a larger $\Delta\rho/{\rho_{\rm trans}}$ or a stiffer core EOS (i.e., larger $c_s^2$ for HybQS and HybSS, or larger $\gamma$ for IHS)  maps to a more extended slow stable branch. 
\end{itemize}
These features also manifest in a generalized parameter scan presented in the Appendix \ref{extended}.
We expect these features to also apply to conventional slow stable hybrid stars.

Additionally, Fig.~\ref{fig_S1} also shows that HybSSs have some peculiar behaviour compared to other types: for cases of transition near $2\, M_\odot$ (solid curves), they develop hybrid branches that cross the XTE J1914-338's $M$-$R$ region with $\partial M/\partial P_c>0$, meaning that they are even stable in the rapid transition context, while those with $\partial M/\partial P_C<0$ are usually termed ``disconnected branch" in rapid phase transition context~\cite{Alford:2013aca}. Furthermore, without invoking Scenario 2, they can already satisfy GW190814's $M_{\rm TOV}>2.59^{+0.08}_{-0.09}\, M_{\odot}$ constraint, provided a late phase transition (dashed curves).


\begin{figure*}[htb!]
\centering
\includegraphics[width=8cm]{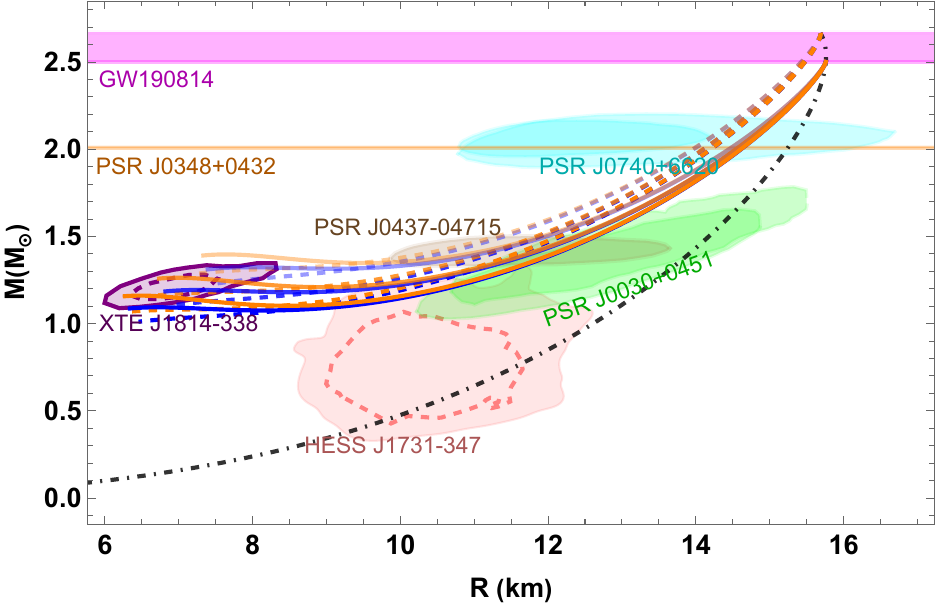}   
\includegraphics[width=8cm]{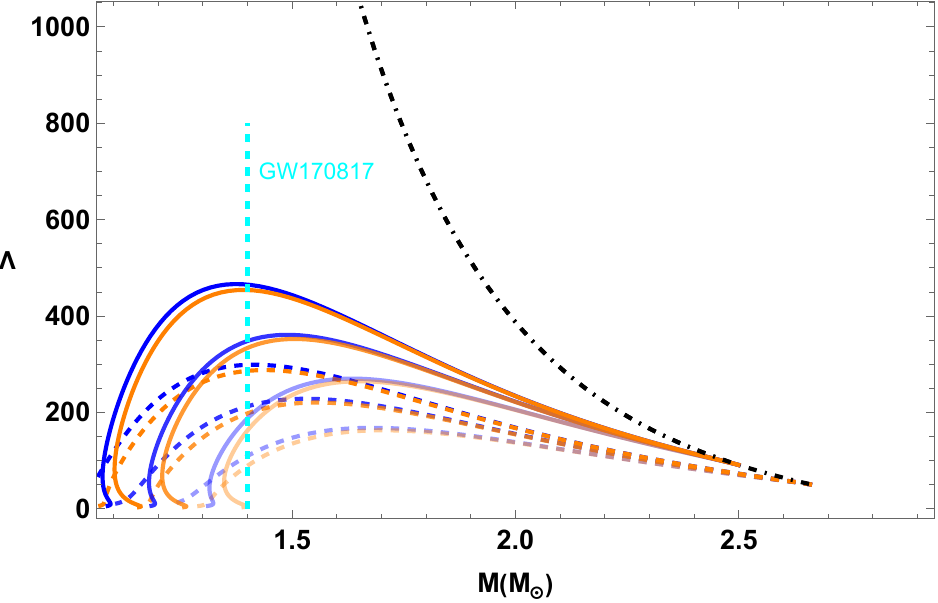}   
\includegraphics[width=8cm]{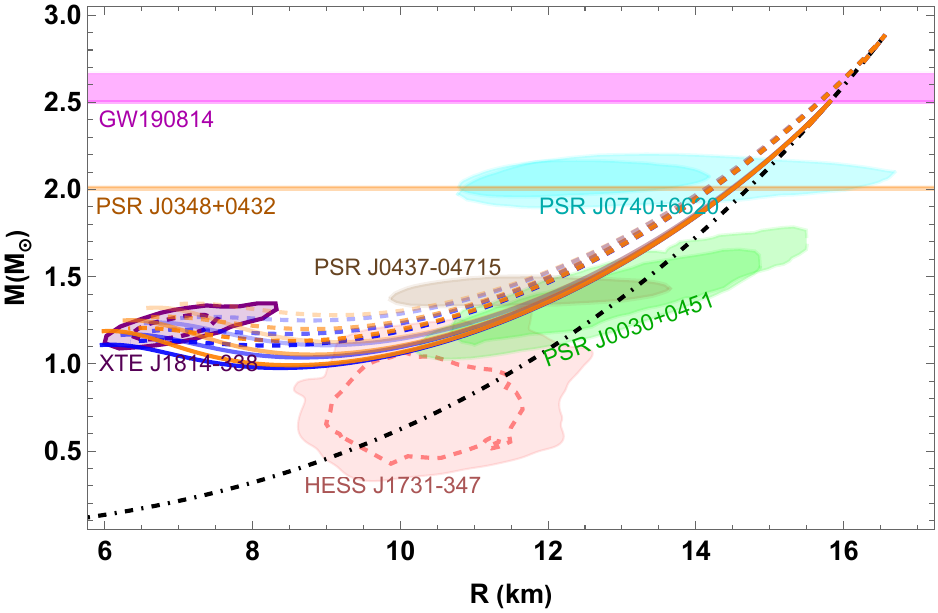}   
\includegraphics[width=8cm]{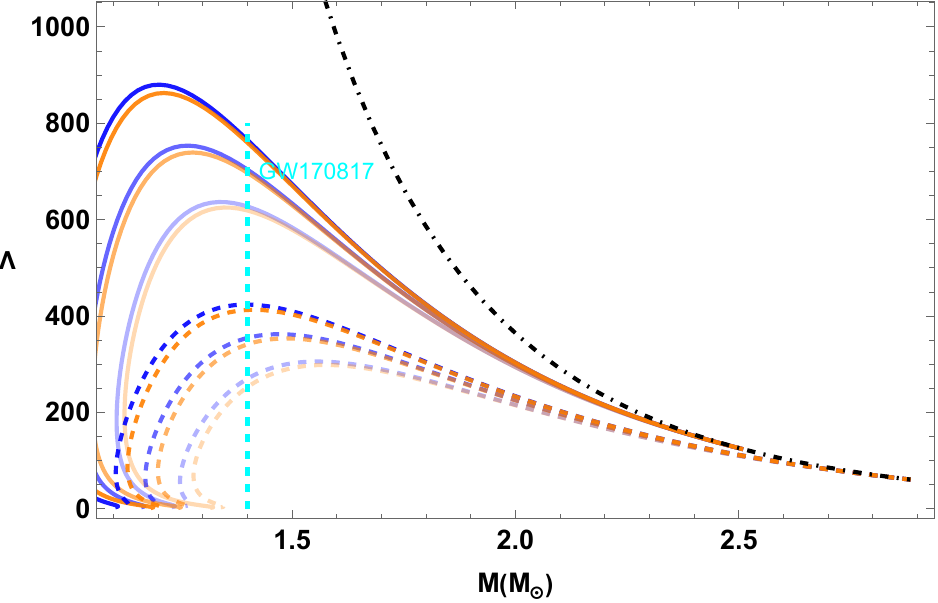}   
 \caption{Mass-radius relations (left) and tidal deformability versus star mass (right) for slow stable HybQSs (first row) and HybSSs (second row), all in Scenario 2 where $\Lambda_{1.4M_\odot}<800$ (GW170817) is only met by the hybrid branch. For the first row, $B=30\rm \, MeV/fm^3$ for the crust quark matter.  
For the second row, $\epsilon/N_q=200/9$ MeV for the crust strangeon matter, fixing $n_s=0.3\,\rm fm^{-3}$.  Blue ($c_s^2=0.8$) and orange ($c_s^2=1$) curves darken with larger $\Delta \rho/\rho_{\rm trans}$, sampling $(4,5,6)$ and $(8,9,10)$ for HybQSs and HybSSs, respectively. For all rows, solid and dashed line styles denote cases where the transition point is close but above the lower mass bound ($\sim 2.5 M_\odot$) of the GW190814 constraint, where $P_{\rm trans}=59\rm\, MeV/fm^3$ and $83\rm\, MeV/fm^3$ for HybQSs, $P_{\rm trans}=39 \rm\, MeV/fm^3$ and $48\rm\, MeV/fm^3$ for HybSSs, respectively.   All curves are terminated at end points where $\omega_0^2= 0$. Black dot-dashed curves are nonhybrid branches. Color bands show astrophysical constraints.} 
\label{fig_S2}
\end{figure*}

For Scenario 2 which addresses GW170817's tidal deformability constraints only with the hybrid branch,  we directly choose large $\Delta \rho/\rho_{\rm trans}$, stiff EOS and large $P_{\rm trans}$ to get a more extended slow stable branch, saving us efforts of tuning parameters, benefiting from the general features observed in Scenario 1 as discussed above. 
We display the results in Fig.~\ref{fig_S2}, skipping showing the drawing of $\omega_0^2(P)$ since they share a similar feature as Scenario 1,  but instead provide the results of tidal deformability to show the main difference. We can easily identify the viable solutions for HybQSs and HybSSs that satisfy all the constraints, even including the GW190814, benefited from a very stiff nonhybrid branch with a late sharp phase transition, leaving the extended hybrid branch satisfying the tidal deformability bounds of GW170817. Note that some HybQS examples and all the HybSS examples cross the XTE J1914-338's $M$-$R$ region with $\partial M/\partial P_c>0$, thus are even stable in the rapid transition context like Scenario 1. The very recent PSR J0437-4715 constraint is easily met by the slow stable branches of all star types. However, by varying the simple polytrope model and even its piecewised form (i.e., piecewise treatment of another one or two polytrope EOSs of different parameter set at higher densities) for the hadronic EOS, we find that the slow stable branches of IHSs that overlaps with GW190814's large mass region can hardly reach the XTE J1814-338's $M$-$R$ region, mainly because the polytrope model of their core hadronic matter would easily violate causality ($c_s^2>1$) in order to make its EOS stiff enough to meet GW190814 while at even much higher center pressure have a long extended stable slow branch to reach XTE J1814-338's $M$-$R$ region. So we neglect this case in this figure.


The viable configurations that cover the XTE J1814-338 region have generally small tidal deformabilities due to their small radius, which may bring concerns against some existing constraints on lower bounds of tidal deformabilities from various analyses on the GW170817 event. On one hand, this problem can be relieved by noting that their hybrid twin counterparts (same mass but larger radius) have generally much larger tidal deformabilities, as can be seen in the right columns of Fig.~\ref{fig_S2} and Figs.~\ref{extendend_B30}-\ref{extendend_B25}. On the other hand,  those analyses that obtained lower bounds of the tidal deformability are mostly targeting neutron matter rather than the self-bound type of matter and the ultra-strong phase transition we explore here. For example, Ref.~\cite{LIGOScientific:2018cki} obtained $\Lambda_{1.4M_{\odot}}=190^{+390}_{-120}$, via polytropic parameterization of nuclear matter EOS and EOS-insensitive relations such as the one associated with tidal deformability $\Lambda$ and compactness $C$ $(\equiv M/R)$, however, neither applies to the self-bound matter case such as quark matter since they require different parameterization frameworks and do have totally different  EOS-insensitive relations on $\Lambda$-$C$~\cite{Chan:2015iou, Yagi:2016bkt}. In another more general study, Ref.~\cite{LIGOScientific:2018hze} obtained $\tilde{\Lambda}=300^{+420}_{-230}$ (with chirp mass $M_c=1.186^{+0.01}_{-0.01} M_\odot$ and mass asymmetry $q=$0.73--1.00)  utilizing the EOS-insensitive relation between spin-induced quadrupole moment ($Q$) and tidal deformability. However, this $Q$-$\Lambda$ relation is expected to be violated by the ultra-strong phase transitions ($\Delta \rho/\rho_{\rm trans}>1$), as Refs.~\cite{annala2018holographic,han2019tidal,lin2019universal,hoyos2022holographic} indicated that this relation can be violated at a $20\%$ level. Despite their imperfect applicabilities, a such lower bound $\tilde{\Lambda}\gtrsim 70$ would necessarily imply $\Lambda_{1.362 M_{\odot}}\gtrsim 70$ assuming chirp mass $M_c=1.186 M_\odot$ and symmetric binary mass ($q=1$). We can see that this lower bound can be easily satisfied by the slow stable twin branches (configurations with $\partial M/\partial P_C<0$ with some being twin counterparts of those in XTE J1814-338 region) in Fig.~\ref{fig_S2} and Figs.~\ref{extendend_B30}-\ref{extendend_B25}.

Larger values of lower bounds on tidal deformabilities come from kilonova analyses of GW180817/AT2017gfo. Ref.~\cite{Radice:2017lry} obtained $\tilde{\Lambda}\gtrsim 400$ as required to eject matter heavier than 0.05 $M_{\odot}$ which the authors assumed to be essential to reproduce the high luminosity of AT2017gfo, while later this lower bound is lowered to 300~\cite{radice2019multimessenger}, 279~\cite{coughlin2019multimessenger}, 242~\cite{Kiuchi:2019lls} with closer analyses of AT2017gfo. In particular, Ref.~\cite{Kiuchi:2019lls} showed that if the maximum mass is large enough to avoid immediate postmerger collapse to a black hole, then the ejecta mass can be substantial to meet AT2017gfo irrespective of the binary tidal deformability, so they expect that even a smaller value of the lower bound can be achieved if the maximum mass can be larger than $2.1 M_\odot$. Therefore, besides the difference in their applied neutron matter EOSs compared to our self-bound matter, the ultra-large maximum mass ($\gtrsim 2.5 M_\odot$) we achieved in Scenario 2 is expected to decrease the lower bound of binary tidal deformability. A reasonable expectation thus might be at the $\tilde{\Lambda}\gtrsim 200$ level, which would agree with Ref.~\cite{bauswein2019equation} despite different motivations. This lower bound of $\tilde{\Lambda}$ implies  $\Lambda_{1.362\, M_\odot}\gtrsim 200$ assuming chirp mass $M_c=1.186 M_\odot$ and symmetric binary mass ($q=1$), which would potentially exclude branches with very large $P_{\rm trans}$ or/and relatively small $\Delta\rho/\rho_{\rm trans}$ considering the resulting relatively small radii, such as the lightest-color dashed and solid curves in the top-right panel of Fig.~\ref{fig_S2} (i.e., HybQS with $\Delta\rho/\rho_{\rm trans}=4$). Therefore, more strict and general results on the tidal deformability lower bounds in the future can potentially help constrain or exclude this new type of objects.  

\section{Summary and Outlook}
Recent astrophysical observations challenge conventional neutron star models: compact stars identified in HESS~J1731-347 ($M \approx 0.77\, M_\odot$, $R \approx 10.4\, \rm km$) and XTE~J1814-338 ($M \approx 1.21\, M_\odot$, $R \approx 7.0\, \rm km$) exhibit anomalously small radii, while GW190814's secondary component ($M \approx 2.6\, M_\odot$) significantly exceeds typical neutron star masses without violating GW170817 constraints from various analyses. In this work, we resolve these tensions by proposing a novel class of self-bound hybrid stars featuring a sharp but slow first-order phase transition, i.e., a large $\Delta\rho/\rho_{\rm trans}$ but with phase transition timescales that exceed oscillation periods. A larger energy density jump ($\Delta\rho/\rho_{\rm trans}$) or stiffer core EOSs extend this slow stable branch to smaller radii, enabling explanations of ultracompact objects. 

Our key findings demonstrate that self-bound hybrid stars such as HybQSs, IHSs, and HybSSs all accommodate HESS~J1731-347 and XTE J1814-338 together with other observational constraints via their slow stable branches, and even via the rapid stable branches for HybSSs. In general, the self-bound feature (i.e., mass grows with radius as center pressure increases at the small-mass regime) benefits saturating HESS~J1731-347, while the extended stable hybrid branch benefits meeting XTE J1814-338.
We also showed the advantage of slow stable self-bound hybrid stars in explaining GW190814's anomalously massive object by allowing a very stiff EOS of the nonhybrid branch if leaving the slow stable hybrid branches to meet the GW170817 constraint, which turns out to be very feasible for HybQSs and HybSSs. Therefore, our work has demonstrated the first working example that reconciles all the astrophysical tensions related to compact stars' masses, radii, and tidal deformabilities, revealing new intriguing possibilities for the behaviour of matter and stars under extreme conditions. 

While the IHS case turns out to have difficulties in meeting GW190814 and XTE J1814-338 at the same time using the simple $e$-type polytrope model and its piecewised version for its core hadronic matter EOS, the chance is still open for other EOS constructions of hadronic matter, such as $\mu$-type polytrope~\cite{Read:2008iy}, more generalized polytope constructions~\cite{OBoyle:2020qvf}, spectral representation~\cite{Lindblom:2010bb}, or the speed-of-sound parameterization~\cite{Greif:2018njt}. It is also more exhaustive to explore a general Bayesian analysis of all parameter space on this new subject. We leave these for future studies. 

There are potential distinct signatures of slow stable self-bound hybrid stars compared to other types of compact stars. The gravitational-wave asteroseismology of this new object may also exhibit interesting features~\cite{Zhang:2023zth}, particularly their $g$-mode of nonradial oscillations~\cite{Sun:2025zpj,Guha:2024gfe,Zhao:2025pgx}, which may have significantly large frequencies~\cite{Tonetto:2020bie}. In addition, the unusually large density discontinuities at their phase transition interface will likely cause a large violation of neutron star universal relations~\cite{han2019tidal,hoyos2022holographic,annala2018holographic,Ranea-Sandoval:2022izm,Ranea-:2023ixr}. Moreover, the strong phase transition is accompanied by a release of energy from the changes of internal energy and gravitational potential energy, potentially sourcing supergiant glitches~\cite{Ma:1996vv}, core-quakes~\cite{Bejger:2005am} and electromagnetic wave radiations such as gamma-ray burst~\cite{Loeb_1998,Menezes:2006zx} and fast radio burst~\cite{shen2023neutron,Wang:2024opz}, which may also originate from the self-bound star's surface~\cite{Geng:2021apl,Xu:2025wwb}.

\begin{acknowledgments}
\subsection*{Acknowledgments}

We thank Chun Huang and Zhi-qiang Miao for useful discussions. We also thank the anonymous referee for helpful suggestions on the improvement of this work. C.Z is supported by the Fundamental Research Funds for the
Central Universities and the Jockey Club Institute for Advanced Study at The Hong Kong University of Science and Technology. J.M.Z.P acknowledges support from ``Fundação Carlos Chagas Filho de Amparo à Pesquisa do Estado do Rio de Janeiro'' -- FAPERJ, Process No. SEI-260003/000308/2024. R.-X.X is supported by the National SKA Program of China (No.2020SKA0120100).
\end{acknowledgments}

\setcounter{equation}{0}
\setcounter{figure}{0}
\setcounter{table}{0}
\setcounter{page}{1}
\makeatletter
\renewcommand{\thefigure}{A\arabic{figure}}

\bigskip

\section{Appendix: Extensive Parameter Scan}
\label{extended}
Here, we explore a more general parameter scan for Scenario 2, considering that it helps reconcile the tension
between the gravitational wave events
GW190814 and other observations. We take the hybrid quark star case, where we have four parameters ($B$, $P_{\rm trans}$, $\Delta\rho/\rho_{\rm trans}$, $c_s^2$). We vary $B$, $P_{\rm trans}$ in reasonable levels (within the $E/A$ stability constraint and the large mass constraint, respectively), and explore the very expanded parameter space of $\Delta\rho/\rho_{\rm trans}$ from $1$ to $10$, and $c_s^2$ from a very small value $c_s^2=0.1$ to the causal limit $c_s^2=1$. The result is shown in the following Figs.~\ref{extendend_B30}-\ref{extendend_B25}, with all curves truncated at zero $\omega_0^2$ of radial oscillations to guarantee their stability in the slow phase transition context, among which those configurations satisfying $\partial M/\partial P_C>0$ are also stable in the rapid phase transition context. 

We can see the general trend that a larger $\Delta\rho/\rho_{\rm trans}$ (darker color depth) and a larger $c_s^2$ extend the slow stable branches more. However, a larger $c_s^2$ lifts the extended branches to higher masses so that decreases the tidal deformability due to increasing compactness, being opposite to the effect of a larger $\Delta\rho/\rho_{\rm trans}$. Therefore, requiring an extended stable hybrid branch will exclude small $\Delta\rho/\rho_{\rm trans}$ and small $c_s^2$, while an upper or lower bound on tidal deformability will constrain $\Delta\rho/\rho_{\rm trans}$ and $c_s^2$ in opposite directions.

We can see that there is a sizable parameter space that allows all observations to be satisfied at the same time. As argued above, the results favor relatively large $\Delta\rho/\rho_{\rm trans}$ and large $c_s^2$. Besides, from Figs.~\ref{extendend_B30_pLow} and~\ref{extendend_B25}, we can see that with lower $P_{\rm trans} $ or lower $B$, some branches that cross the XTE J1814-338 region have mass increases as the center pressure increases. This means they are stable even in the rapid transition context. 

\begin{figure*}
\centering

\begin{minipage}{\textwidth}
\centering
\includegraphics[width=0.3\textwidth]{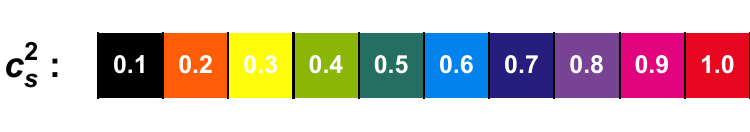} 
\end{minipage}
\includegraphics[width=7.4cm]{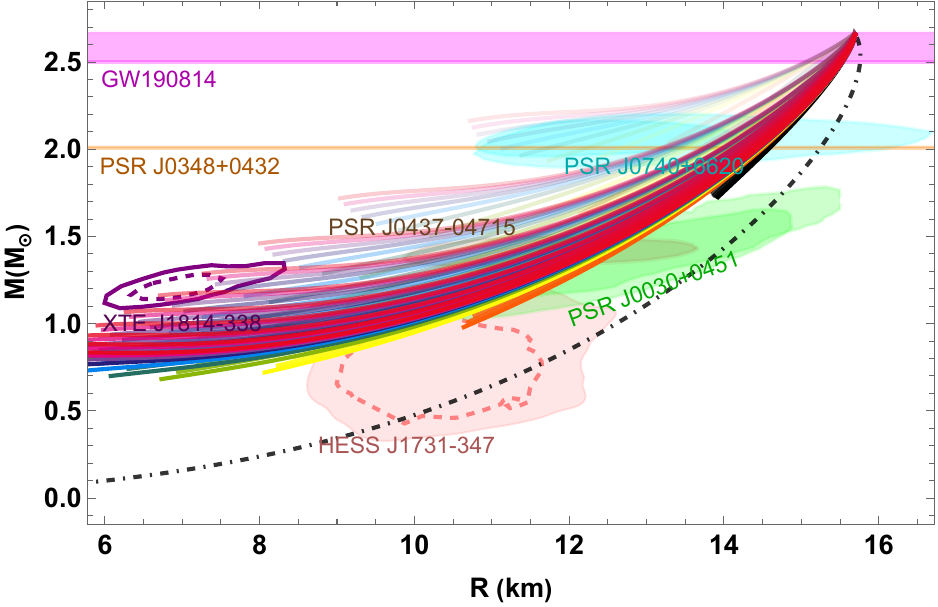} 
\includegraphics[width=7.4cm]{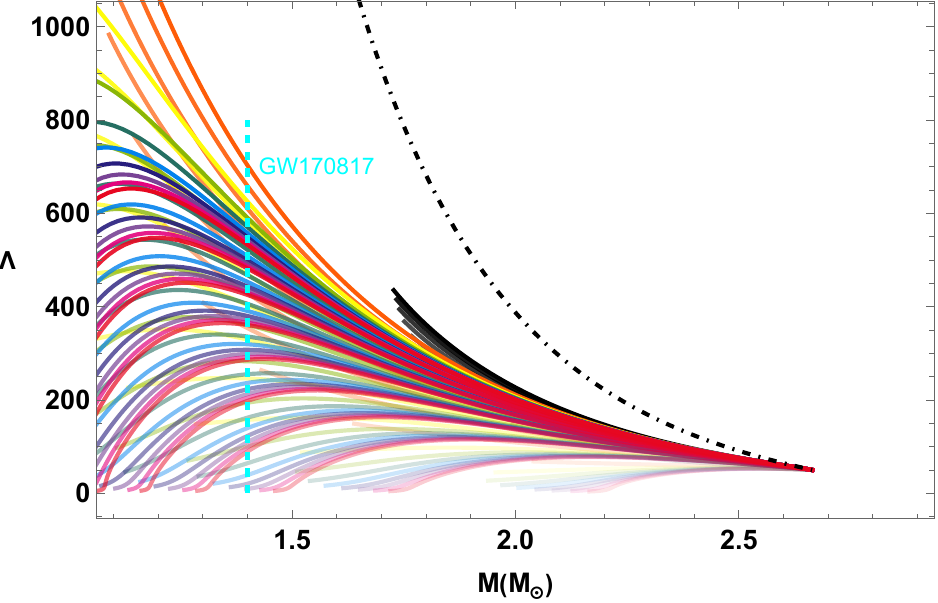} 
\includegraphics[width=7.4cm]{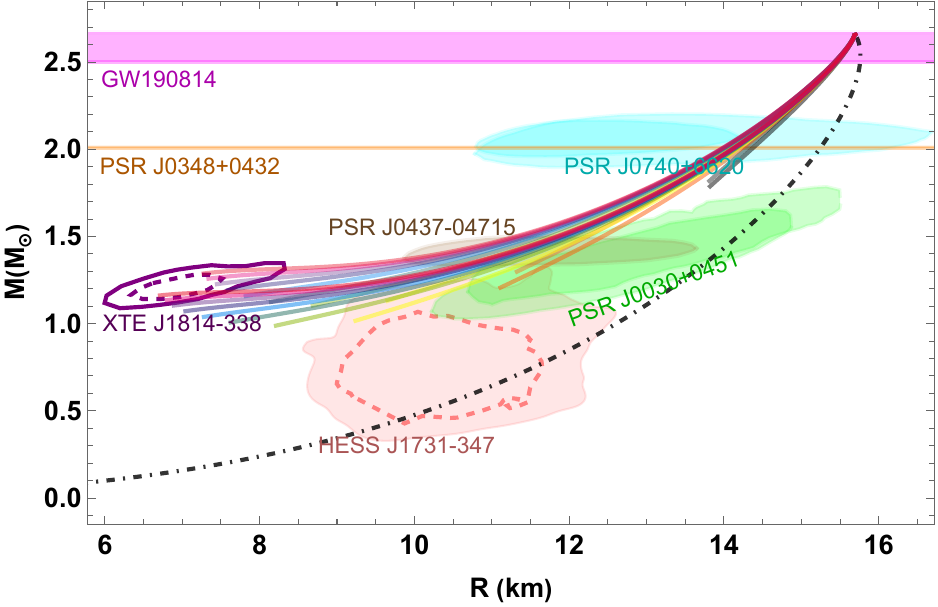}   
\includegraphics[width=7.4cm]{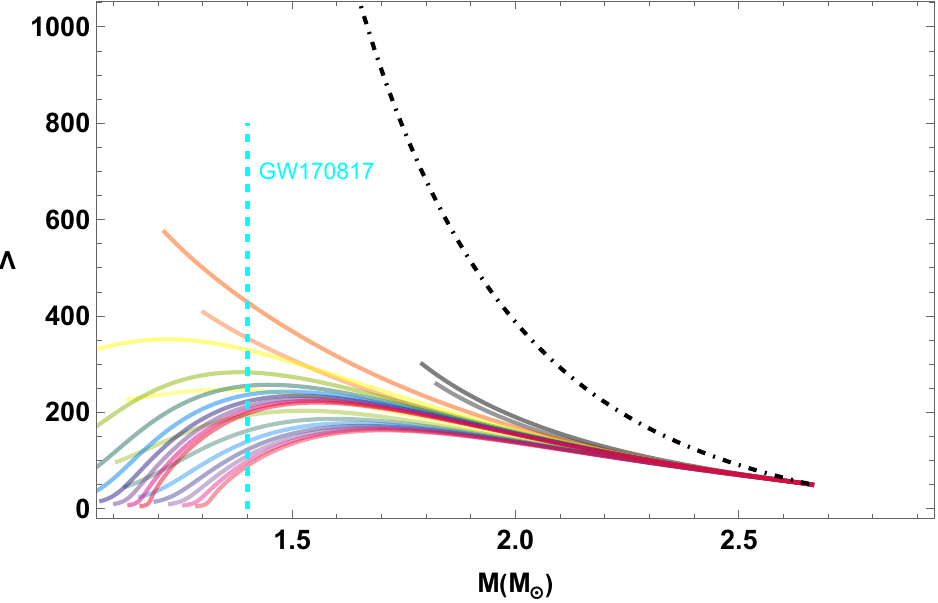}   
 \caption{$M$-$R$ (left) and $\Lambda$-$M$ (right) relations for slow stable HybQSs of $B=30\rm\, MeV/fm^3$ with $P_{\rm trans}=83\rm\, MeV/fm^3$, featuring a expanded $\Delta\rho/\rho_{\rm trans}=(1,2,...,10)$ (first row) from lighter to darker depth of each color, and a more constrained range $\Delta\rho/\rho_{\rm trans}=(4,5)$ (second row) that can have viable $c_s^2$ that satisfy XTE J1814-338. All curves are truncated at the $\omega_0^2=0$ point.
}
\label{extendend_B30}
\end{figure*}

\begin{figure*}
\centering
\includegraphics[width=7.4cm]{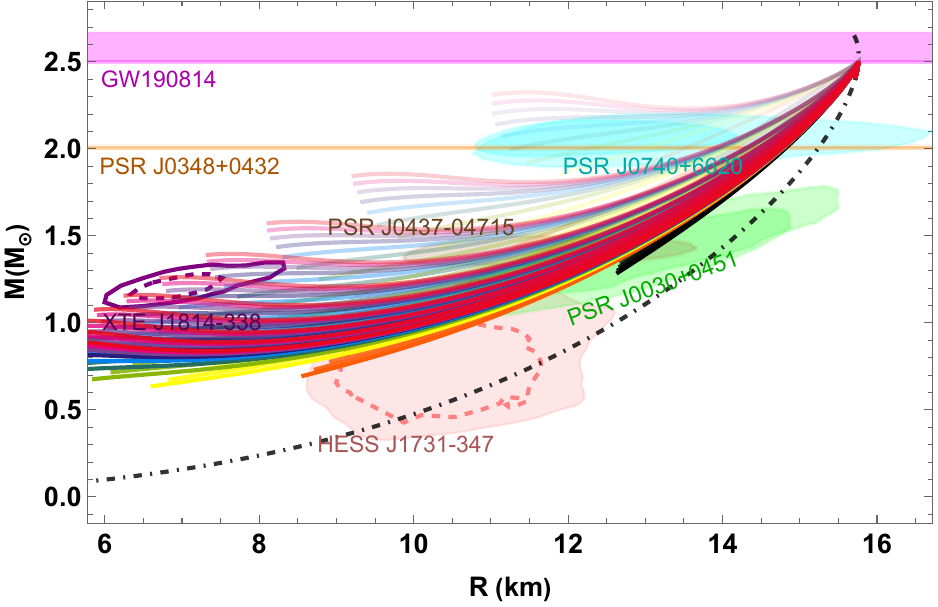} 
\includegraphics[width=7.4cm]{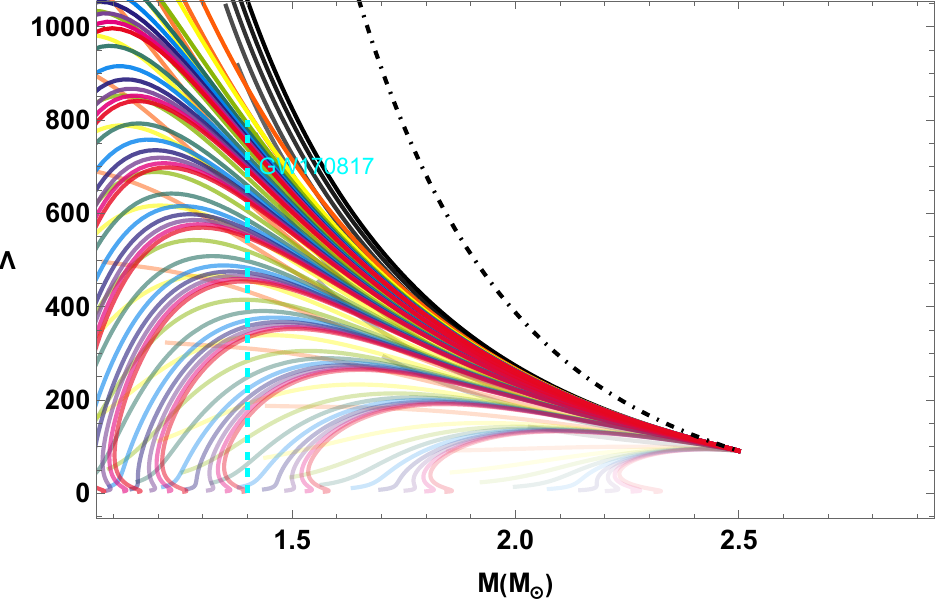} 
\includegraphics[width=7.4cm]{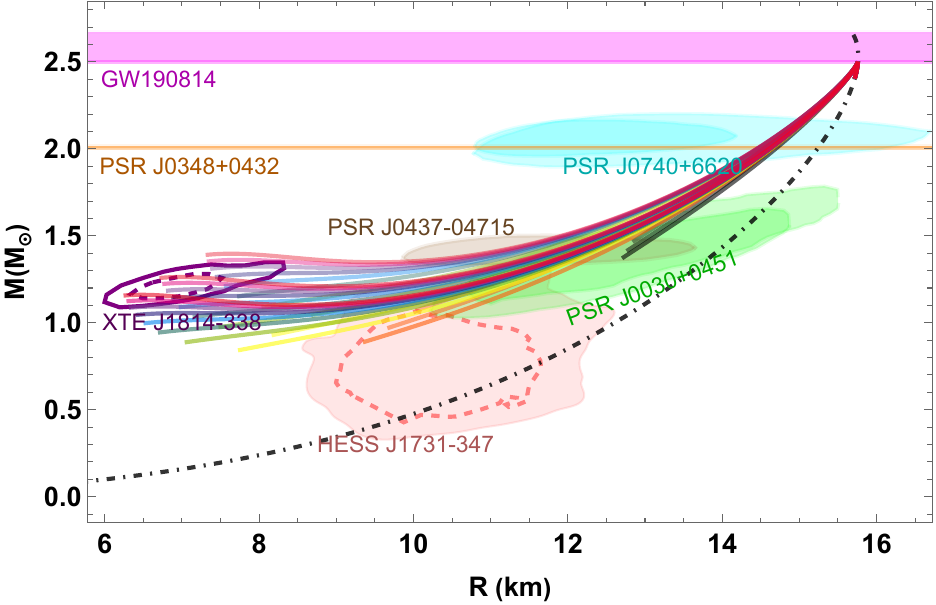}
\includegraphics[width=7.4cm]{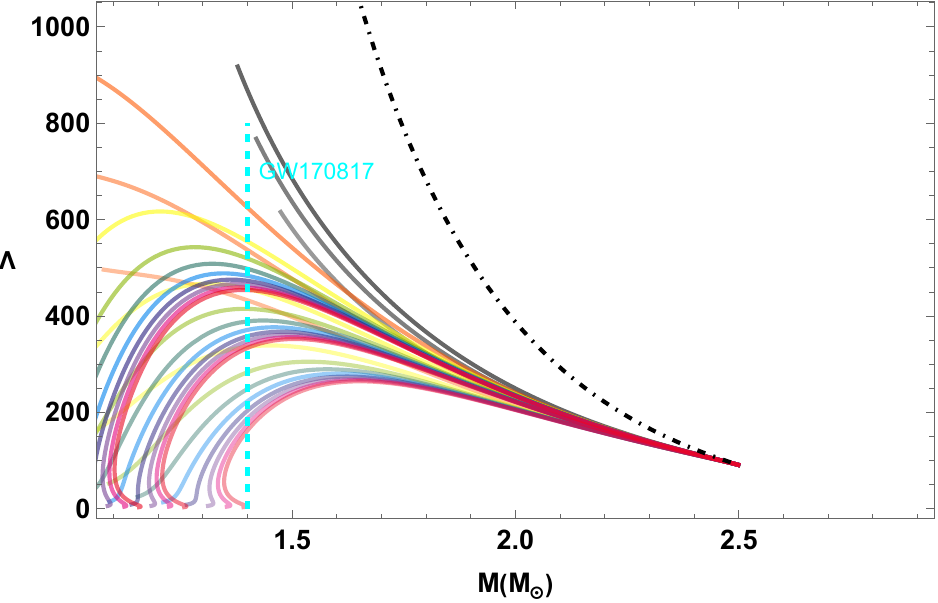}
 \caption{Similar to Fig.~\ref{extendend_B30} but with smaller $P_{\rm trans}=59\rm\, MeV/fm^3$. 
The range constrained by satisfying XTE J1814-338 region (second row) becomes $\Delta\rho/\rho_{\rm trans}=(4,5,6)$.
}
\label{extendend_B30_pLow}
\end{figure*}

\begin{figure*}[htb]
\centering
\begin{minipage}{\textwidth}
\centering
\includegraphics[width=0.3\textwidth]{figs/barl.pdf} 
\end{minipage}
\includegraphics[width=7.5cm]{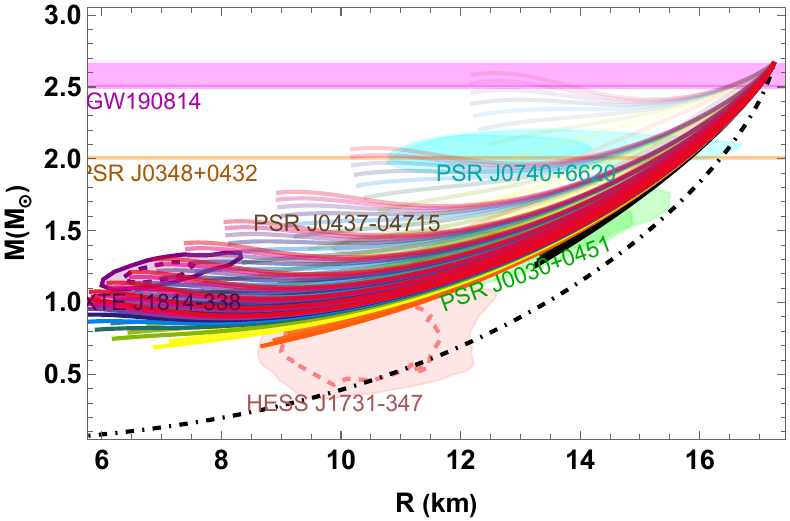}  
\includegraphics[width=7.5cm]{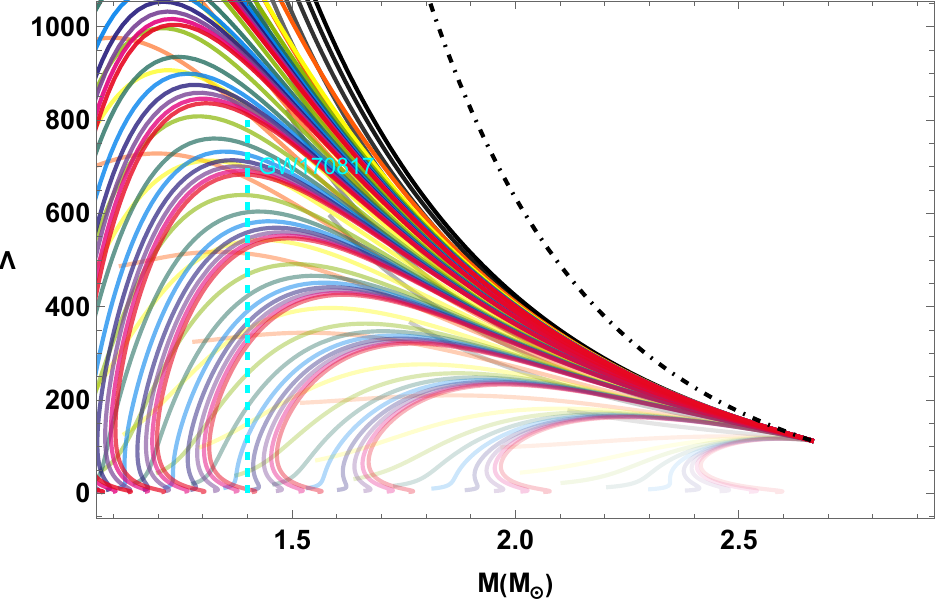}  
\includegraphics[width=7.5cm]{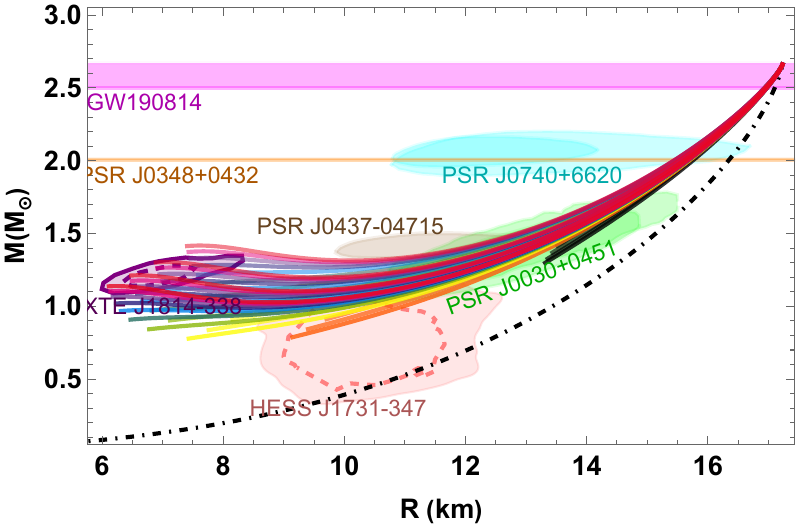}   
\includegraphics[width=7.5cm]{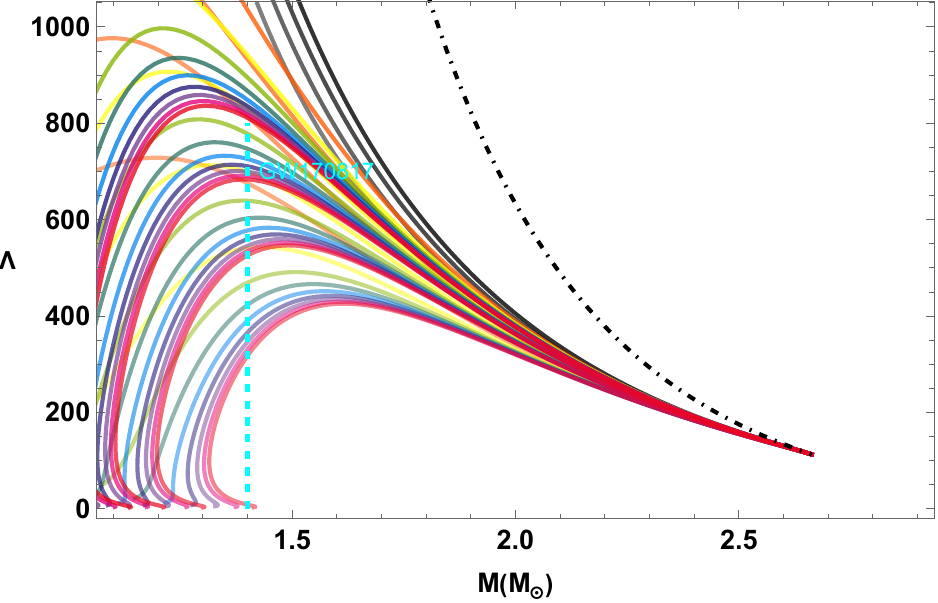}   
 \caption{Similar to Fig.~\ref{extendend_B30} but with smaller $B=25\rm\, MeV/fm^3$ with $P_{\rm trans}=44\rm\, MeV/fm^3$ (so that the transition occurs at the same top mass bound of GW190814). The XTE J1813-338 constrained range (second row) becomes $\Delta\rho/\rho_{\rm trans}=(5,6,7,8)$.
}
\label{extendend_B25}
\end{figure*}




\clearpage
\bibliography{slow}
\end{document}